\documentclass[%
 aip,
 amsmath,amssymb,
 reprint,%
]{revtex4-1}

\usepackage{graphicx}% Include figure files
\usepackage{dcolumn}% Align table columns on decimal point
\usepackage{bm}% bold math
\usepackage{mathrsfs}
\usepackage[utf8]{inputenc}
\usepackage[T1]{fontenc}
\usepackage{etoolbox}
\usepackage{bm}
\usepackage{xfrac}
\usepackage{xcolor}

\makeatletter
\def\@email#1#2{%
 \endgroup
 \patchcmd{\titleblock@produce}
  {\frontmatter@RRAPformat}
  {\frontmatter@RRAPformat{\produce@RRAP{*#1\href{mailto:#2}{#2}}}\frontmatter@RRAPformat}
  {}{}
}%
\makeatother

\begin{document}

\preprint{AIP/123-QED}

\title{Reconfiguration of Amazon's Connectivity in the Climate System}% Force line breaks with \\
\author{Adam Giammarese}%
\altaffiliation{Email: amg2889@rit.edu}
%\email{amg2889@rit.edu}
\affiliation{School of Mathematical Sciences, Rochester Institute of Technology, Rochester, NY 14623, USA}
\author{Jacob Brown}
\affiliation{Department of Mathematics, University of Connecticut, Storrs, CT 06269, USA}

\author{Nishant Malik}
\altaffiliation{Email: nxmsma@rit.edu}
%\email{nxmsma@rit.edu}
\affiliation{School of Mathematical Sciences, Rochester Institute of Technology, Rochester, NY 14623, USA}

\begin{abstract}
With the recent increase in deforestation, forest fires, and regional temperatures, the concerns around the rapid and complete collapse of the Amazon rainforest ecosystem have heightened. The thresholds of deforestation and the temperature increase required for such a catastrophic event are still uncertain. However, our analysis presented here shows that signatures of changing Amazon are already apparent in historical climate data sets. Here, we extend the methods of climate network analysis and apply them to study the temporal evolution of the connectivity between the Amazon rainforest and the global climate system. We observe that the Amazon rainforest is losing short-range connectivity and gaining more long-range connections, indicating shifts in regional-scale processes. Using embeddings inspired by manifold learning, we show that Amazon connectivity patterns have become more variable in the twenty-first century. By investigating edge-based network metrics on similar regions to the Amazon we see the changing properties of the Amazon are significant in comparison. Furthermore, we simulate diffusion and random walks on these networks and observe a faster spread of perturbations from the Amazon in recent decades. Our methodology innovations can act as a template for examining the spatiotemporal patterns of regional climate change and its impact on global climate using the toolbox of climate network analysis. 
\end{abstract}

\pacs{}% insert suggested PACS numbers in braces on next line

\maketitle %\maketitle must follow title, authors, abstract and \pacs

\begin{quotation}
The Amazon rainforest is an ecological system of high social significance, identified as a tipping element of the global climate system---increasing global temperatures and deforestation could lead to its dieback. Such an event will have disastrous consequences for the local environment and communities living in the region. Furthermore, the Amazon rainforest is an important carbon sink, and its destruction will negatively impact the planet's climate. Amazon is also facing an immediate threat due to the highest level of deforestation in history, which is already having adverse impacts on global and local environmental, weather, and climate patterns. This work attempts to understand changes occurring in patterns of interactions between the global climate system and the Amazon over the last seven decades. Traditionally, such a study will require broad analysis for large-scale computational models; however, here, we take an entirely data-driven approach and identify the changes occurring in the connectivity between the global climate and the Amazon rainforest through climate network analysis. Our study shows that the connectivity between the Amazon rainforest and the global climate system has been experiencing reconfiguration in their connectivity patterns. 
\end{quotation}

\section{Introduction}
\label{sec:intro}

\begin{figure*}
    \centering
    \includegraphics[width=0.9\linewidth,keepaspectratio]{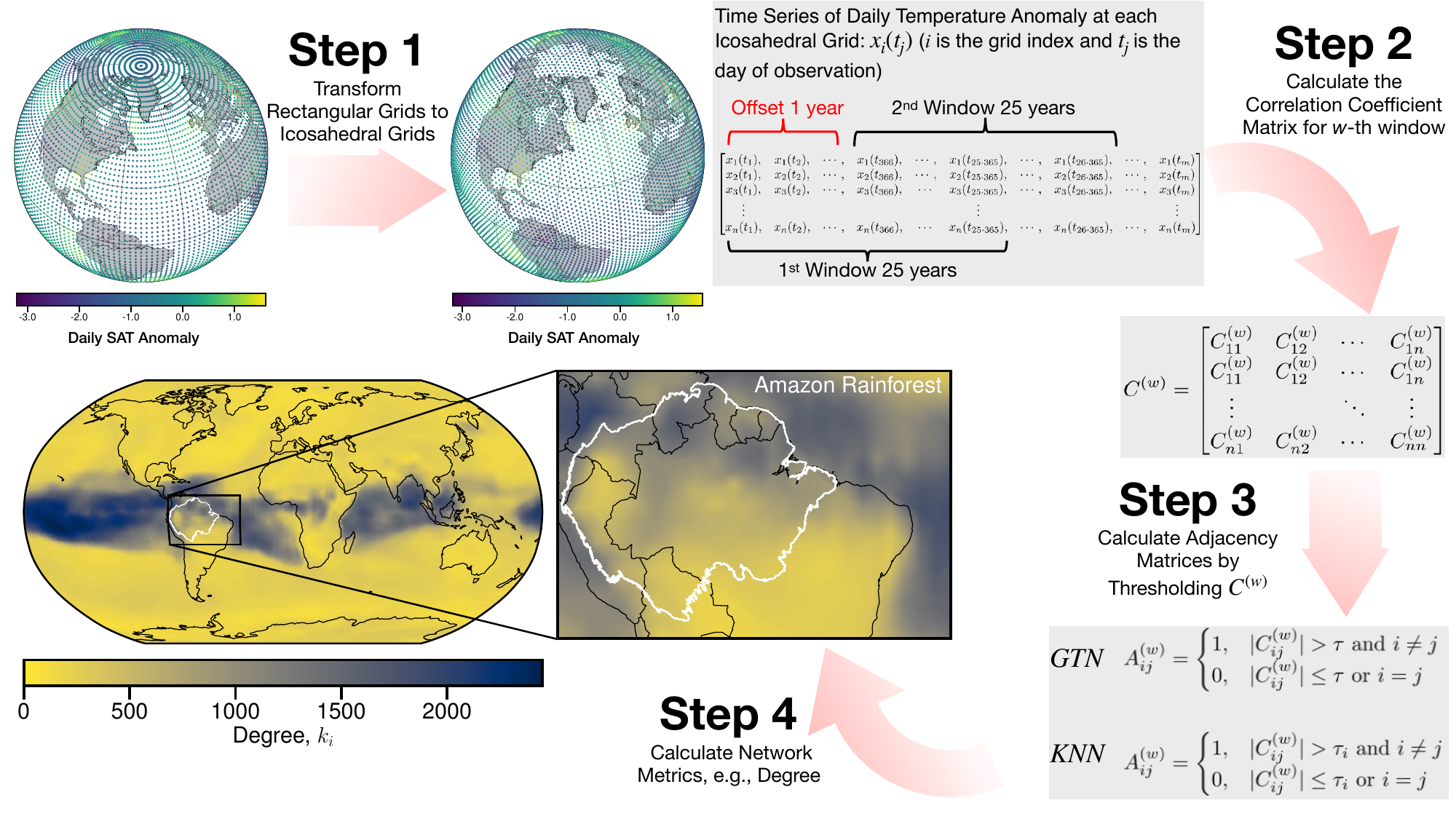}
    \caption{This diagram illustrates various steps involved in our analysis by constructing networks from the NCEP/NCAR Reanalysis 1 surface air temperatures (SAT) data set. The map shows a sample outcome of our analysis, where we show the $k_i$ (degree) in one of the resulting networks. }
    \label{fig:dia1}
\end{figure*}

Humanity faces an existential crisis in the form of climate change.\cite{RN23, biosci-biz088} Across the globe, communities are suffering due to increasing temperatures, rising sea levels, and the growing intensity of extreme weather phenomena such as floods, droughts, and hurricanes.\cite{pnas181139, biosci-biz088,lenton2019climate,RN23,012420-105026,Church2011,2018GL081183} %The destruction of ecosystems due to climate change also acts to exacerbate climatic changes themselves, advancing a cascade of harmful environmental events. 
Climate change is also driving the destruction of ecosystems, and simultaneously collapsing ecosystems are further exacerbating the climate crisis.\cite{gcb15539, conl12348}
The Amazon rainforest is one such ecosystem, also classified as one of the tipping elements of the global climate system; it may dieback if global temperature increases by $3-4^\circ$C, which will have catastrophic consequences for the region in the form of changing precipitation patterns and aridification. Currently, the Amazon rainforest is under even more immediate threat---historically highest levels of deforestation.\cite{boers2017deforestation, Boulton2022} It is understood that if one-fourths of the forest is lost, it could lead to a chain of events that will also result in dieback of the forest.\cite{boers2017deforestation, Boulton2022,lenton2019climate} The destruction of the Amazon rainforest will have far-reaching consequences for the planet, as it plays a critical role in controlling the global carbon dioxide fluxes, the most significant greenhouse gas. Amazon rain forest also modulates the local and global rainfall patterns through evapotranspiration.\cite{avissar2005global}

Estimates from the end of 2019 show that over 718,000 {km$^2$} of rainforest have been deforested since 1970.\cite{butler} Recently, it has also been hypothesized that Amazon may have passed a critical point where the unabated deforestation has turned it into a net carbon dioxide source.\cite{Gatti2021} Given this precarious situation of the Amazon, it is paramount to study further the consequences of changes in the Amazon on the global climate system, including developing mathematical and computational tools that can provide physical insights into these changes. The traditional approach to such a study will require a large-scale computational modeling effort, simulating a hierarchy of interactions between various climate sub-systems and phenomena. In contrast, we propose a simpler, data-driven approach known as climate network analysis, which uses existing historical data sets and employs the toolbox of network science to study the evolution of connectivity between the global climate system and the Amazon. Using this technique, we analyze large spatiotemporal surface air temperature data for the last seven decades and show a reconfiguration of interactions between Amazon and the global climate system.

%Within climate physics, the question of how interactions between the Amazon and the global climate are evolving has not been well studied. Traditionally, one would have to carry out a large number of computationally expensive simulations of general circulation models. In contrast, we utilize computationally simpler methods developed primarily by the physics community to study climate, nowadays, classified as climate network analysis. 

The underlying assumption in climate network analysis is that the global climate system is a complex network of numerous phenomena manifesting on various spatial and temporal scales, where the nodes in these climate networks are geographic locations, and the edges represent interactions between various phenomena.\cite{Malik2012,Ozturk2019,Liu2023,tsonis2006networks,Donges2009,radebach2013disentangling,Boers2019} For example, it is well known that monsoon systems interact with El Niño–Southern Oscillation (ENSO), North Atlantic Oscillation (NAO), and Indian Ocean Dipole (IOD), and climate networks at the global scales do tend to have edges representative of these interactions.\cite{tsonis2006networks,Donges2009,radebach2013disentangling,Boers2019} The primary advantage of working with climate networks is that they provide a simpler mathematical representation of information contained in massive spatiotemporal climate datasets, and in the last decade, several significant insights into the global and regional climates have been obtained through climate network analysis. \cite{Malik2012,Boers2019,Liu2023,radebach2013disentangling}  Although less explored, climate networks are excellent tools for studying changes in weather patterns and climate due to anthropogenic forcings, as these forcings get engraved into the structure of the climate and can be identified and explored using the methods of network analysis. Thus, we will use the climate network as an analogous form to study climatic phenomena in relation to the Amazon rainforest. Furthermore, we will implement tools previously not explored in the climate networks settings, such as manifold learning, diffusion, and random walks on graphs.  

A limitation of our study that requires to be underscored here is that we do not claim that the deforestation of the Amazon is causing the changes highlighted between the Amazon and the global climate system. Our study can not establish such causal links. The changes we are reporting are caused by the interplay of many factors, including the region's environmental degradation and other variations occurring on the regional and planetary scale in the climate system due to increasing temperatures. We cannot discern the factors causing the reconfiguration reported in this study. Nonetheless, the Amazon rainforest is a critical component of the global climate system; it influences the global energy balance, the hydrological cycle, and the carbon balance.\cite{avissar2005global} Therefore, quantifying changes occurring in the structure and patterns of interactions between the Amazon and the global climate will improve our understanding of climate change and the role of Amazon in it.

\section{Materials and Methods} 
\label{sec:m-and-m}

\subsection{Dataset}
\label{sec:dataset}

We construct networks using surface air temperature data from the NCEP/NCAR Reanalysis 1 data set provided by NOAA's (National Oceanic and Atmospheric Administration's) Physical Sciences Laboratory (PSL).\cite{data-ncep} This data results from cooperation between NCEP (National Centers for Environmental Prediction) and NCAR (National Center for Atmospheric Research).\cite{TheNCEPNCAR40YearReanalysisProject} We choose the average daily air temperature at the $\sigma=0.995$ level (the height where air pressure is $99.5\%$ of surface air pressure) for our analysis, as the surface air temperature (SAT) is one of the most accurately and extensively measured climate variables and an excellent proxy for evolving dynamics of the global climate system.\cite{ipcc-2014,https2008GL037155} This data spans from 1948 to 2022 and are structured on a $2.5^\circ$ by $2.5^\circ$ regular latitude-longitude grid, for a total of $10,512$ grid points. 

\begin{figure*}
    \centering
    \includegraphics[width=0.9\linewidth,keepaspectratio]{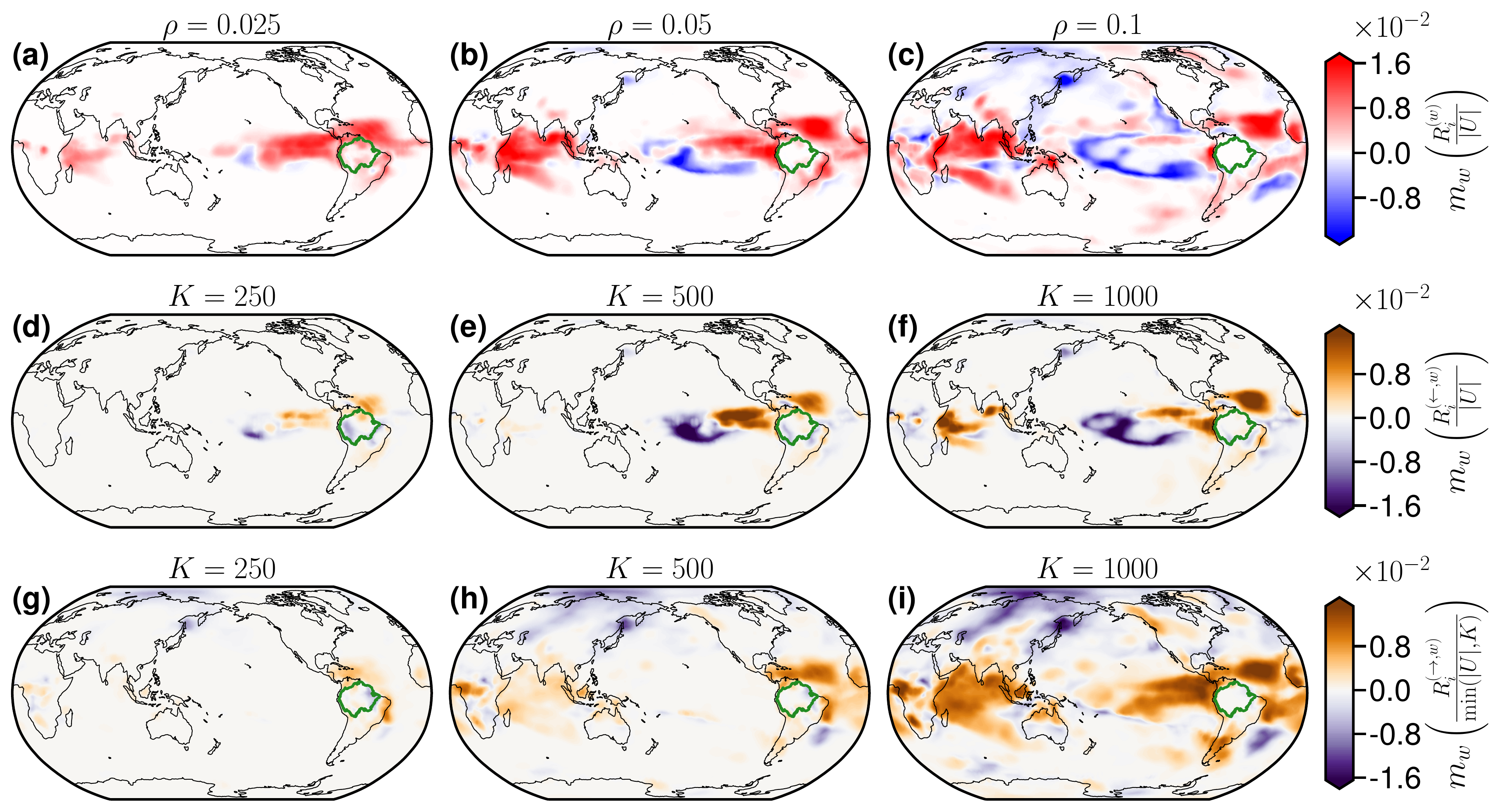}
    \caption{Trends in the local connectivity to (and from) Amazon in global threshold networks (GTN) and K-nearest neighbors networks (KNN) as $\rho$ (global density of links) and $K$ (number of outgoing edges for each node) is varied, respectively. $m_w$ is the slope of the trend in the density of connections between Amazon and a node $i$, i.e., $R_i^{(w)}/{|U|}$.  Observe that as $\rho$ and $K$ increase, changes in the long-range connections start to appear while the changes in local connections within Amazon become stable. The Amazon is gaining connectivity in the tropics, especially the tropical Atlantic and the eastern Pacific, and the Indian Ocean region, some of the most dynamic regions in the global climate system. Simultaneously, Amazon is losing links in the Central Pacific as well. Note that in KNN graphs an edge $i\rightarrow j$ exists when $j$ is a K-nearest neighbor of $i$, and therefore panels (d,e,f) show when node $i$ is in the neighborhood of the Amazon, and panels (g,h,i) show when the Amazon is in the neighborhood of node $i$.}
    \label{fig:Tr1Amazon}
\end{figure*}
% \begin{figure*}
%     \centering
%     \includegraphics[width=0.9\linewidth,keepaspectratio]{AmazonConnectivityTrendIn(Varyk).pdf}
%     \caption{Trend in Amazon connectivity at $\rho=0.05$.}
%     \label{fig:Amazon}
% \end{figure*}

\subsection{Network Construction and Analysis}
\label{sec:nc-and-a}
Our analysis involves four steps, out of which the first three transform the above dataset into networks. The fourth step is the analysis of these networks using various network analysis techniques, including analysis based on a variety of network metrics. Below we describe these four steps; the diagram in Fig.~\ref{fig:dia1} provides an illustration summarizing these steps. \\

%\noindent {\bfseries{\emph{Step 1: Regridding to Icosahedral Grid}}} \\
\subsubsection*{Step 1: Regridding to Icosahedral Grid}
%\label{sec:step1}

A drawback of the regular latitude-longitude grid is that the density of grid points is greater near the poles than at the equator, leading to spatial biases when calculating network properties. One technique to counter this spatial bias in the network metrics is to take into account the area of the grid squares created by the latitude and longitude lines, which are smaller near the poles than at the equator, resulting in a method known as area-weighted connectivity.\cite{tsonis2006networks} An alternative technique we also employ here is to project the data onto a grid pattern free of such spatial biases. We use an icosahedral grid constructed in the following manner: first, an icosahedron with edge length two is identified in the Euclidean space. Each edge is bisected, a vertex is added at each bisection point, and an edge is added between each pair of adjacent points (i.e., those pairs separated by distance one). This process is repeated five times to give a total of $10,242$ vertices on the surface of an icosahedron. Note that at the $n$-th iteration, edges are added between vertices separated by a distance of $1/2^{n-1}$. The vertices are then projected onto a sphere of radius one, and the Cartesian coordinates of the vertices are converted to latitudes and longitudes. The final output is a set of $10,242$ latitude-longitude pairs that are evenly spaced around the globe. After the projection, the nodes on the icosahedral grid are assigned a temperature time series via bilinear interpolation. 

\subsubsection*{Step 2: Calculate Correlation Matrix}
%\label{sec:step2}

After re-gridding the data, we calculate correlations between grid points, where we first obtain the surface air temperature anomalies (SATA) for each node by subtracting long-term daily average temperatures from the SAT time series and normalizing it by the standard deviation of SAT time series. This process is equivalent to taking the z-score of SAT time series at each node, and it eliminates the long-term mean annual cycle from the data, resulting in SATA time series of length $m$ at node $i$: $\{x_i(t_j)\}_{j=1}^m$, where $t_j$ represents the time index. We divided the period January, 1948 to December, 2022 into 51 time windows, with each time window being 25 years long and an offset of 1 year between each time series (365.25 days). The time scale was chosen so that the number of El Ni\~no cycles was the same for all networks. We also created a single network for the entire time span of the data set. One 25-year window will consist of $m=25 \times 365.25 \approx 9131$ time points. For every $w$-th time window, we calculate the Pearson correlation coefficient $C_{ij}^{(w)}$ between grid points $i$ and $j$, these coefficients can be arranged into an $N \times N$ matrix  ${\bf{C}}^{(w)}=[C_{ij}^{(w)}]_{N \times N}$ where $C_{ij}^{(w)} \in [-1,1]$ and $N=10242$ is the total number of grid points.

\subsubsection*{Step 3: Calculate Adjacency Matrix (Construct Networks)}
%\label{sec:step3}

We construct two types of networks from the correlation matrices: global threshold networks (GTN) and k-nearest neighbor networks (KNN).  \\

{\emph{Global Threshold Networks (GTN)}: An adjacency matrix represents a network or a graph. Here we obtain $w$-th adjacency matrix ${\bf{A}}^{(w)}$ corresponding to the $w$-th time window by thresholding ${\bf{C}}^{(w)}$ as follows:  $A_{ij}^{(w)}=1$ if  $|C_{ij}^{(w)}|>\tau$ and $i \neq j$ (meaning there exists an edge $i \leftrightarrow j$), else $A_{ij}^{(w)}=0$.  The threshold $\tau$ is so chosen that the network has a $\rho$ link density. This choice of threshold is based on our limited focus on studying the most persistent features of the atmospheric and climatic processes, encoded into the strongest correlations in the data. This step transforms large spatiotemporal data into a sparse binary matrix; that is it converts a complex and large dataset into simpler mathematical objects, a graph, or network $G^{(w)}$ represented by the binary adjacency matrix ${\bf{A}}^{(w)}$. 

{\emph{K-Nearest Neighbor Networks (KNN)}: The k-nearest neighbors (KNN) is a non-parametric unsupervised learning framework used in clustering, classification, and regression of data; however, here, we use KNN to generate networks from SATA data. For a given node $i$ we take the $K$ other nodes $j\neq i$ such that $|C_{ij}^{(w)}|$ are amongst the $K$ greatest entries in $|C_{i*}^{(w)}|$, excluding $|C_{ii}^{(w)}|$; one may also hold the view that for each node $i$ we find an individual threshold $\tau_i$ such that there are $K$ indices $j\neq i$ such that $|C_{ij}^{(w)}|\geq \tau_i$. For each node $j$ amongst the greatest $K$ correlation values, a directed edge $i\rightarrow j$ is added in the graph, indicating that $j$ is in the nearest neighbors of $i$. While it remains true that the global edge density remains fixed in KNN graphs (more particularly, there will always be $K\cdot N$ edges), it differs from GTN due to its selection of local correlation thresholds for each node versus a global threshold.

%We created $48$ networks, each with a time scale of $25$ years, with a $1$ year offset between each network. The time scale was chosen so that the number of El Ni\~no cycles was the same for all networks. We also created a single network for the entire time span of the data set. To specifically delineate the Amazon region, we created a box of $166$ nodes over the Amazon rainforest, spanning 5$^{\circ}$N - 17$^{\circ}$S latitude and 50$^{\circ}$W - 80$^{\circ}$W longitude. See the diagram in Fig.~\ref{fig:dia1} for a succinct description of the above-described network construction process. 

\subsubsection*{Step 4: Network Metrics and Analysis}
%\label{sec:step4}

%Our network construction process culminates in $48$ networks: $\{G^{(l)} \}_{l=1}^{l=48}$, $t_l$ covers $l-th$ time window consisting of $25$ consecutive years of the studied period. 

Step 3 concludes the network construction process, resulting in a set of $51$ adjacency matrices $\{{\bf{A}}^{(w)}\}_{w=0}^{w=50}$, each of these adjacency matrices represents a particular network $G^{(w)}$. The set of networks  $\{G^{(w)} \}_{w=0}^{w=50}$ corresponds to the $51$ time windows with a $25-$year span and offset of $1$ year covering the period from January, $1948$ to December, $2022$. In this last step, we analyze these resulting networks identifying emerging features in Amazon's interaction with the global climate system. Next, we introduce our analysis's mathematical notations, metrics, and methods.

 \subsubsection{Relative Connectivity of the Amazon} \label{sec:r-connect}
 
 The total number of nodes in any given network is $N=10,242$, which is the same as the total number of grid points resulting after regridding. The nodes in the Amazon region (the white outlined region in the maps in Fig.~\ref{fig:dia1}) form the set $U$ with cardinality $|U|=166$. $R_i^{(w)}$ is the number of links between the node $i$ and the Amazon in the GTN network $G_{GTN}^{(w)}$, while $R_{i}^{(\leftarrow,w)}$ is the number of links from the Amazon to node $i$ and $R_{i}^{(\rightarrow,w)}$ is the number of links from the node $i$ to the Amazon in KNN network $G_{KNN}^{(w)}$.

\subsubsection{Trends in Connectivity to (and from) the Amazon}
\label{sec: trends-connect}

 As stated above $\displaystyle R_{i}^{(w)}$ is the density of connections that node $i$ has to the Amazon region, and a simple measure of connectivity shift of the Amazon will be the slope of the linear trend in  $\displaystyle R_{i}^{(w)}$ for every grid point. We denote $m_w(\cdot)$ to be the slope of the linear trend-line over the range of valid $w\in\{0,1,\dots,50\}$ for the respective property.

\subsubsection{Laplacian based Embedding of Amazon nodes:}
\label{sec: lap-b-embd}

A popular embedding for graphs is the eigenvectors of the Laplacian, which forms the basis for various clustering and manifold learning algorithms such as the Laplacian Eigenmaps.\cite{LEBelkin} These manifold learning algorithms are state-of-the-art tools for identifying and visualizing low-dimensional structures in high-dimensional data, and here we use Laplacian Eigenmaps to visualize structural changes in the subgraph of the Amazon with respect to the global climate system.  Laplacian defined as the  ${\bf{L}}_{sym}={\bf{D}}-{\bf{A}}_{sym}$, where ${\bf{D}}={\bf{D}}^{(\leftarrow)}+{\bf{D}}^{(\rightarrow)}$ is the degree matrix (where $\leftarrow$ and $\rightarrow$ refer to incoming and outgoing edges, respectively), a diagonal matrix with $D_{ii}$ entry being the degree of node $i$ and ${\bf{A}}_{sym}=({\bf{A}} + {\bf{A}}^T)/2$ is the symmetric adjacency matrix.  We use the symmetric normalized version of the Laplacian, which is defined as  ${\mathcal{L}}_{sym}={\bf{D}}^{-1/2}{\bf{L}}_{sym}{\bf{D}}^{-1/2}$. For each network, we compute the first $\mu+1$ eigenvectors ${\bf{v}}_0,{\bf{v}}_1,\cdots,{\bf{v}}_{\mu}$ of  ${\mathcal{L}}_{sym}$ associated with eigenvalues $0=\lambda_0 \leq \lambda_1 \leq \cdots \leq \lambda_\mu$. Each of these vectors ${\bf{v}}_i$ consist of $N$ elements, ${\bf{v}}_i=[ v_{ij} ]_{j=1}^N$. We drop the eigenvector ${\bf{v}}_0$ corresponding to eigenvalue $0$ and project each node $i$ in ($\mu=3$)-dimensional space with position of $i$ given by ${\boldsymbol{\xi}}_i=[ v_{1i}, v_{2i}, v_{3i}].$  In theory, a visualization of these manifolds for different networks could provide insights into the structural evolution of networks over time; however, the manifolds extracted using this technique are prone to mirroring due to arbitrary direction along each eigenvector; hence, in practice, such visualization of manifolds of successive graphs is not informative. To overcome this issue, we take an alternative approach, focusing only on the scale of these embeddings by taking the 2-norm of $\boldsymbol{\xi}_i$ for the Amazon nodes, $\lVert {\bf{\xi}}_{i \in U} \rVert_2$ and generate distributions of $\lVert {\bf{\xi}}_{i \in U} \rVert_2$ for different time windows $w$. A systematic expansion (contraction) in these distances would indicate an increasing (decreasing) volume of the Amazon sub-manifold, occupying more space in the global climate system, and allowing faster spread of perturbations. In general, expansion (contraction) of the space described by $\lVert {\bf{\xi}}_{i \in U} \rVert_2$  would be indicative of Amazon reconfiguring its connectivity.

% shows the change in the Amazon's dynamics in relation to the global climate. More specifically, shorter average distances are indicative of increased local dynamics, that is, those dynamics that are confined to the Amazon and its immediate surroundings. Longer average distances are indicative of the opposite: increases global dynamics where the Amazon interacts with more distant regions of the globe. 
%Figure~\ref{fig:connectivity} shows the change in the average distance ratio from 1948 to 2020. 18 randomly placed boxes the same size of the Amazon were also analyzed to examine the comparative change in the Amazon's average distance ratio against the network as a whole.

\subsubsection{Trends in Network Metrics across Random Boxes}
\label{sec: random-boxes}

In order to analyze the changing structure of the Amazon compared to the changing structure of the rest of the globe, we measure average network metrics of edges incident to a variety of random areas across the climate network. We first grow random boxes by initializing a latitude-longitude pair, and expanding a box with the same aspect ratio of the box bounding the Amazon region until it contains roughly the same number of nodes as the amazon ($|U|=166$). Since the average degree is much higher for nodes in the tropical region, we also create tropical random boxes with the same process, but requiring the box be contained within the Tropics of Capricorn and Cancer. We will investigate four kinds of network metrics for these boxes: average edge betweenness centrality of edges remaining in the box (labeled ``in'' edges), average edge betweenness centrality of edges that connect a node inside with a node outside the box (labeled ``out'' edges), connectivity ratio, and average geodesic distance of edges connecting an inside and outside node. Edge betweenness centrality,\cite{edge-betweenness} $\bar{B}_e$, measures the proportion of shortest paths in the graph that cross over edge $e$, and is normalized by the number of edges in the graph; $\bar{B}_e$ measures the importance of edge $e$ based on how many lines of efficient information flow require the use of $e$. We define connectivity ratio, $\Gamma$, to be the proportion of incident edges that leave the box; $\Gamma$ is indicative of the amount of teleconnections a box has. Lastly, $\langle d_{g,out}\rangle$ is the measure of average geodesic distance of edges that leave the box, which is calculated using the Haversine formula.\cite{cajori2007history} For all metrics that involve edges leaving the box (namely $\langle\bar{B}_{out}\rangle$, $\Gamma$, and $\langle d_{g,out}\rangle$) have individual variants for incoming and outgoing edge directions ($\leftarrow$ and $\rightarrow$, respectively) for directed KNN graphs.

\begin{figure}
    \centering
    \includegraphics[width=\columnwidth]{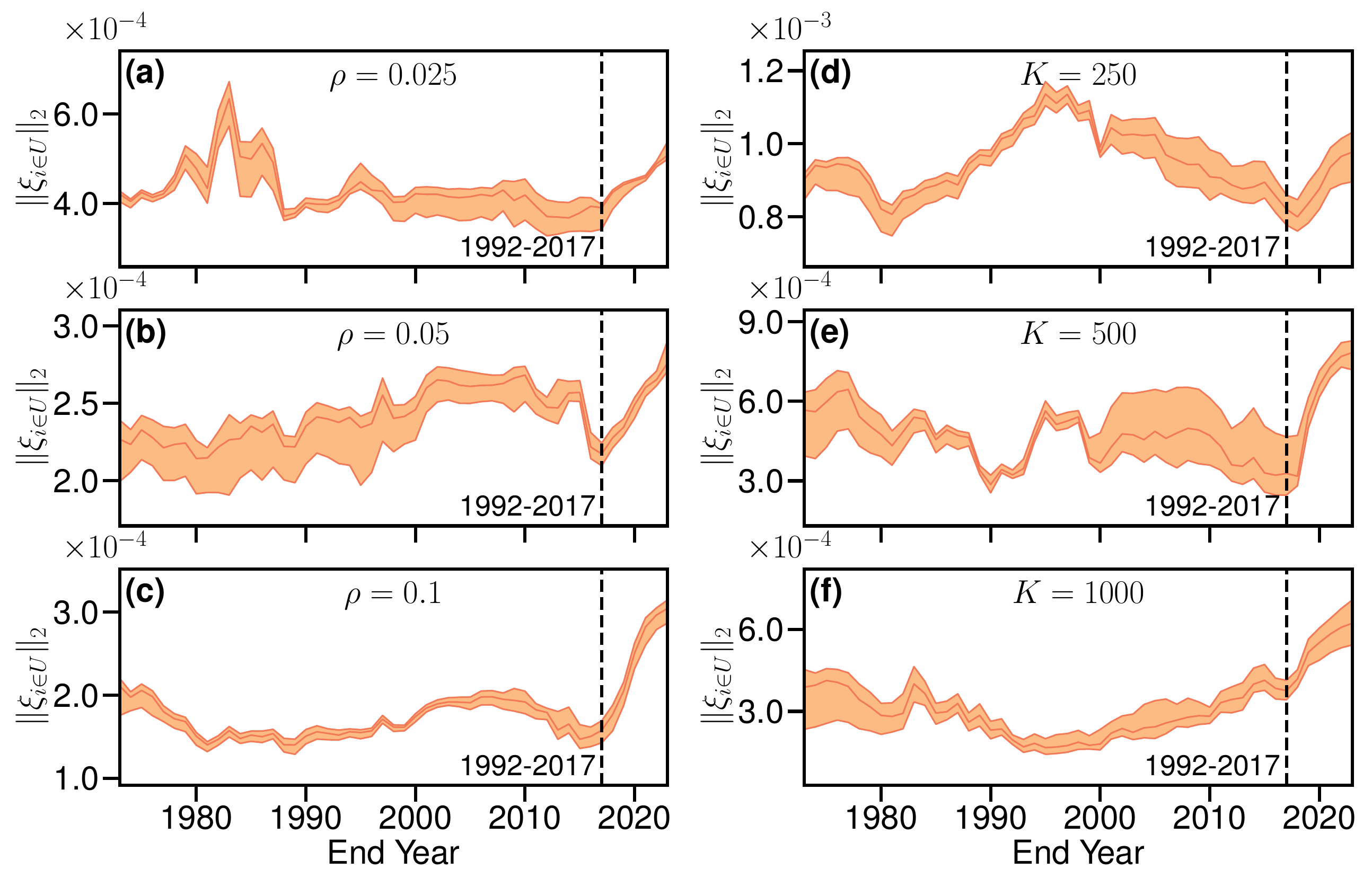}
    \caption{Confidence band (25th percentile to 75th percentile) of $\lVert {\bf{\xi}}_{i \in U} \rVert_{2} $, the 2-norm of each 3D point  ${\boldsymbol{\xi}}_i$ for the Amazon nodes, where the thick line in the middle indicates the median. ${\boldsymbol{\xi}}_i$ are the Laplacian Eigenmap-based projections of the network, see section~\ref{sec: lap-b-embd} for details. After the 1992-2017 time window, we observe an expansion in these norms, indicating the Amazon sub-manifold occupying more space in the global climate manifold,  favoring a connectivity pattern that allows for a faster spread of perturbations from the Amazon.}
    \label{fig:LEdistances}
\end{figure}

\begin{figure}
    \centering
    \includegraphics[width=\columnwidth]{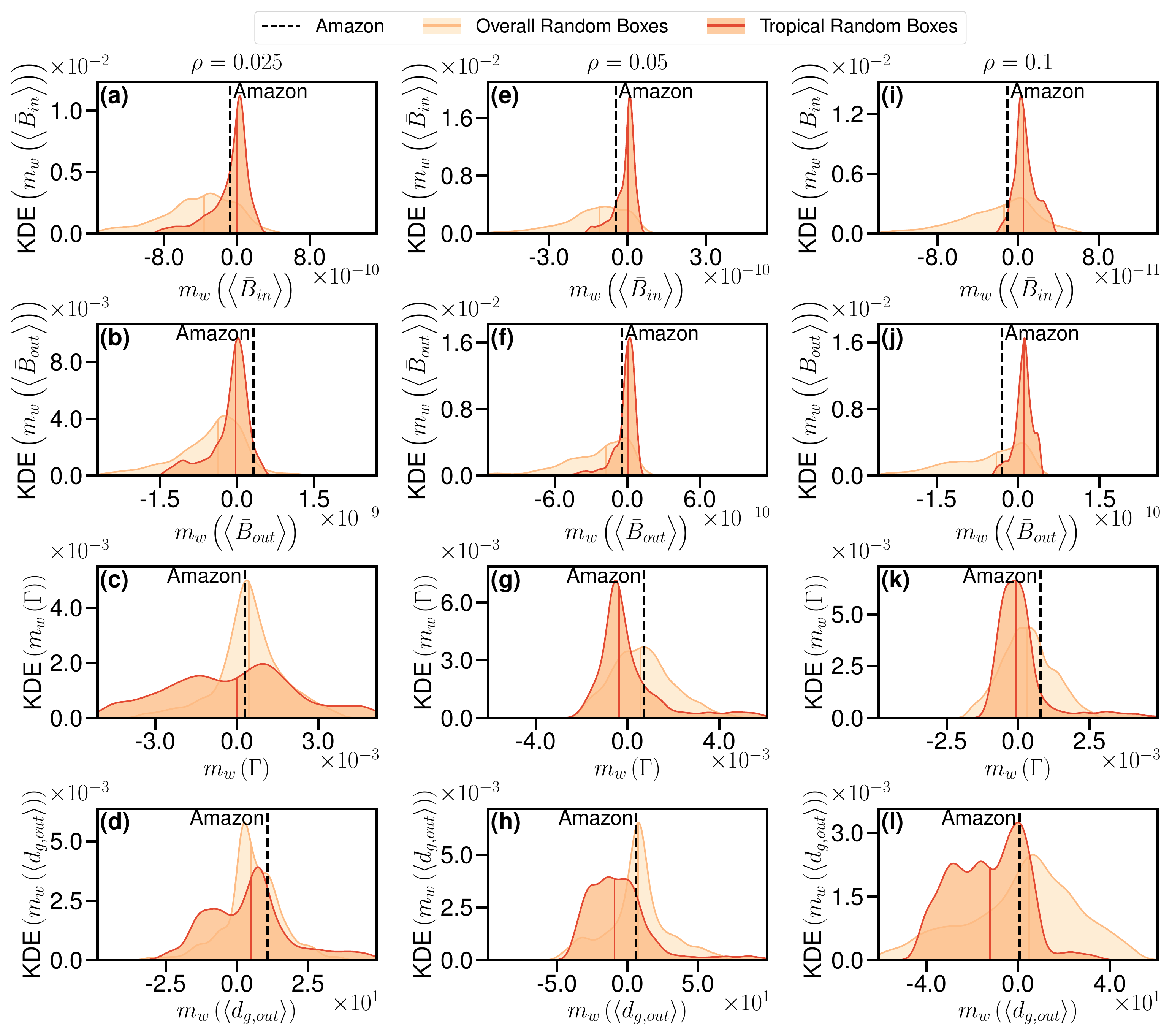}
    \caption{Kernel density estimation (KDE) of trend-line slopes of various edge-based network metrics across 5000 tropical and 5000 overall random boxes for GTN graphs. Panels (a-d), (e-h), and (i-l) are for $\rho=0.025$, $\rho=0.05$, and $\rho=0.1$, respectively; panels (a,e,i) and (b,f,j) display average edge betweenness centrality for edges that remain inside the random box and edges that leave the box, respectively, panels (c,g,k) display the connectivity ratio of the box, and panels (d,h,l) display the average geodesic distance of edges leaving the box. Black lines represent the slope of the Amazon, while the dark and light curves represent the KDE distributions for tropical and overall random boxes, respectively. 5\% of the highest and lowest slope outliers are removed for ease of KDE interpretation, and the vertical lines represent the median of observed slopes. Observe that for many metrics over the values of $\rho$ the slope of the Amazon lies in the extreme of similar tropical boxes, and occasionally in the extreme of overall boxes. See section \ref{sec: random-boxes} for more details on these metrics.}
    \label{fig:met-rhp}
\end{figure}

\begin{figure}
    \centering
    \includegraphics[width=\columnwidth]{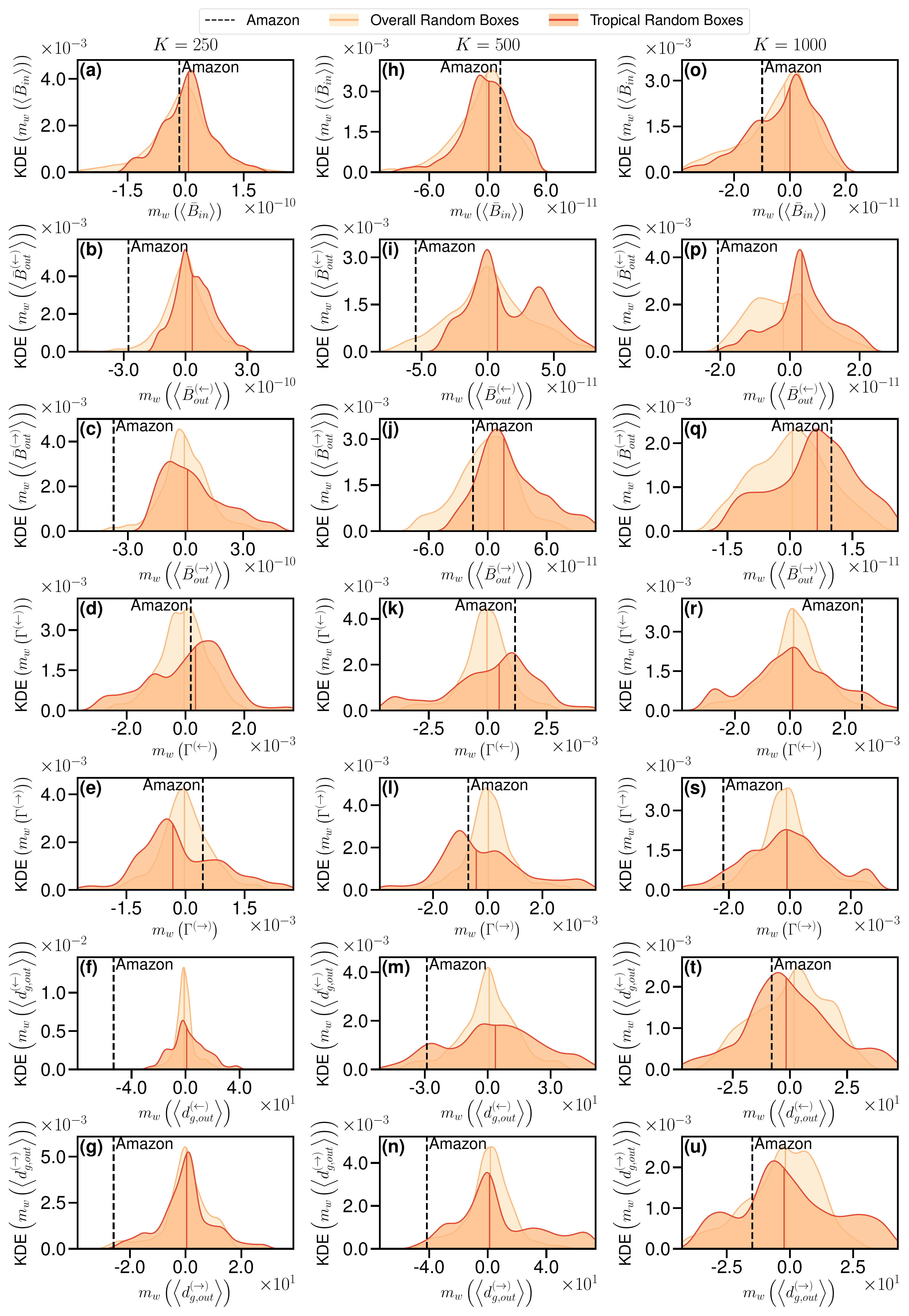}
    \caption{Kernel density estimation (KDE) of trend-line slopes of various edge-based network metrics across 5000 tropical and 5000 overall random boxes for KNN graphs. Panels (a-g), (h-n), and (o-u) are for $K=250$, $K=500$, and $K=1000$, respectively; panels (a,h,o), (b,i,p), and (c,j,q) display average edge betweenness centrality for edges that remain inside the random box and (incoming and outgoing) edges that leave the box, respectively, panels (d,k,r) and (e,l,s) display the (incoming and outgoing) connectivity ratios of the box, and panels (f,m,t) and (g,n,u) display the average geodesic distance of (incoming and outgoing) edges leaving the box. Black lines represent the slope of the Amazon, while the dark and light curves represent the KDE distributions for tropical and overall random boxes, respectively. 5\% of the highest and lowest slope outliers are removed for ease of KDE interpretation, and the vertical lines represent the median of observed slopes. Observe that for many metrics over the values of $K$ the slope of the Amazon lies in the extreme of similar tropical boxes, and occasionally in the extreme of overall boxes. See section \ref{sec: random-boxes} for more details on these metrics.}
    \label{fig:met-k}
\end{figure}

\subsubsection{Diffusion on climate networks from Amazon:}
\label{sec: diffusion}

In order to further investigate the Amazon's changing connectivity with the global climate system, we carry out perturbations in the Amazon using graph diffusion. One may consider graph diffusion to demonstrate how information flows between neighbors. For both the GTN and KNN graphs we build diffusion simulations under the same assumptions: nodes may only send or receive information to their neighbors; nodes may only donate information through outgoing edges to nodes with less information; the rate of information spread between nodes is proportional to their difference in information. We denote the amount of information at node $i$ at time $t$ to be $\phi_i(t)$, $u(\cdot)=(\text{sign}(\cdot)+1)/2$ to be the Heaviside step function, $c$ the coefficient of diffusion (which we choose to be $c=1$), and we use the adjacency matrix convention where $A_{ij}=1$ if there exists an edge $i \rightarrow j$. Note that if the graph is undirected (such as in GTN) Eq.~\ref{diff-general} collapses to the classical undirected graph diffusion equation. Also note that since $\phi_i$ is only donated across outgoing edges to nodes $j$ with less $\phi_j$, diffusion in KNN graphs only spreads to the \textit{nearest neighbors} of node $i$, and is received by nodes $j$ where $i$ is in $j$'s nearest neighbors. In all diffusion simulations, $\phi_{i\in U}(t=0)=N/|U|$ and $\phi_{i\in U^{\mathrm{C}}}(t=0)=0$, and therefore the perturbation over the Amazon will diffuse to the rest of the graph, and since our graphs are all connected all nodes will asymptotically approach $\phi_i(t\to\infty)=1$ due to the conservation of $\sum_{i=1}^N \phi_i(t)$.

\begin{equation}
    \begin{split}
    \frac{d\phi_i}{dt} = c\sum_{j=1}^N \Bigr[ &A_{ij}(\phi_j - \phi_i)u(\phi_i-\phi_j) \\ +  &A_{ji}(\phi_j-\phi_i)u(\phi_j-\phi_i) \Bigr]  \label{diff-general}
\end{split}
\end{equation}

We use RK45 to obtain a numerical solution for Eq.~\ref{diff-general} for each GTN and KNN graph and for each time window. To measure the spread diffusion on the graphs we will use three metrics: average non-Amazon information, proportion of nodes affected by diffusion, and diffusion distance. Average non-Amazon information is calculated as $\langle \phi_{i \notin U}(t)\rangle$, and we expect this value will start at 0 at $t=0$ and asymptotically approach 1; the rate at which this measurement approaches 1 is indicative of how quickly the perturbation diffuses from the Amazon to the rest of the climate system. The proportion of affected nodes is measured simply as the proportion of nodes $i\notin U$ whose $\phi_i(t)$ is above the threshold $|U|/N$: $\sfrac{|\phi_{i\notin U}(t) \geq \sfrac{|U|}{N}|}{|U^{\mathrm{c}}|}$, where $U^{\mathrm{c}}$ is the complement of $U$; we also expect this measurement to start at 0 at $t=0$ and asymptotically approach 1, but its rate is not only indicative of how much information has diffused from the Amazon but also how much of the globe the perturbation has diffused to. Lastly, we define diffusion distance between nodes $i$ and $j$ in time window $w$ as
\begin{equation*}
    d_{i,j}^{(w)}(t) = |\phi_i^{(w)}(t) - \phi_j^{(w)}(t)| \label{diff-dist}
\end{equation*}
which measures a natural sense of distance in the diffusion process: two nodes with equal $\phi$ cannot diffuse to each other (even if they share an edge), while for nodes with differing $\phi$, $d_{i,j}$ is indicative of how much information is left to transfer between the nodes to reach a consensus. To measure an aggregate diffusion distance between node $i$ and the original perturbation, we define
\begin{equation*}
    d_{i,U}^{(w)}(t) = \frac{1}{|U|}\sum_{j\in U} d_{i,j}^{(w)}(t) = \frac{1}{|U|}\sum_{j \in U} |\phi_i^{(w)}(t) - \phi_j^{(w)}(t)| \label{diff-dist-amazon}
\end{equation*}
to be the average diffusion distance between node $i$ and the Amazon.

\subsubsection{Random Walks}
\label{sec: random-walks}

While diffusion restricts the spread of perturbations only to those with less information, we may also investigate the scenario where perturbations spread freely. Thus, we perform random walks originating in the Amazon and observe the regions in which there is changing likelihood of walks being received. We use Markov chain iteration to evolve the likelihood of a walk of length $\eta$ being at node $i$: $\displaystyle p_i(\eta)$. We initialize the probabilities to be $\displaystyle p_{i\in U}(\eta=0)=1/|U|$ and $\displaystyle p_{i\in U^{\mathrm{c}}}(\eta=0)=0$, and we define $\displaystyle \hat{p}_i(\eta) = Np_i(\eta)$ so that $\displaystyle \hat{p}_i(\eta\to\infty)=1$ for connected graphs. The row-normalized transition matrix ${\bf{M}}$ is defined such that when a walk is at node $i$, it has an equal probability of stepping to any neighbors $j$ of $i$ such that there exists an edge $i \rightarrow j$: 
\begin{equation*}
    M_{ij} = \frac{A_{ji}}{\sum_{k=1}^NA_{ki}}.
\end{equation*}
The random walk probability vector is updated as
\begin{equation*}
    {\bf{p}}(\eta) = {\bf{M}}{\bf{p}}(\eta - 1) = {\bf{M}}^\eta {\bf{p}}(0).
\end{equation*}

%\begin{align}
%    M_{ij} &= \frac{A_{ji}}{\sum_{k=1}^NA_{ki}}. \label{transition-matrix}\\
%    p(\eta) &= Mp(\eta - 1) \label{random-walk}
%\end{align}

\begin{figure*}
    \centering
    \includegraphics[width=0.8\linewidth]{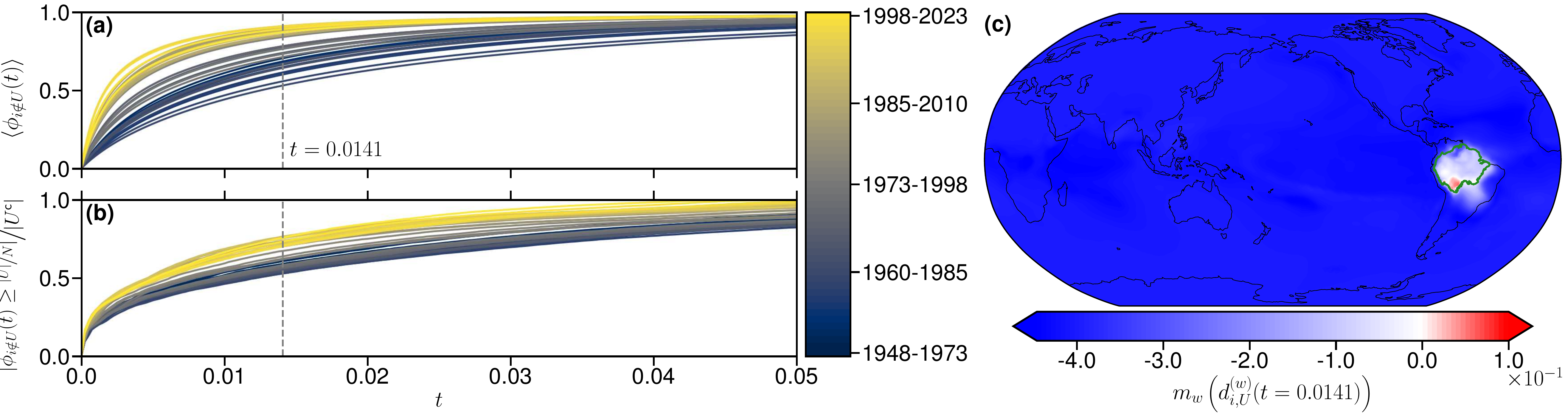}
    \caption{Diffusion simulation results for $\rho=0.025$ GTN graphs. (a) displays the average non-Amazon information, $\langle \phi_{i\notin U}(t)\rangle$, (b) displays the proportion of non-Amazon nodes affected by the diffused perturbation, $\sfrac{|\phi_{i\notin U}(t) \geq \sfrac{|U|}{N}|}{|U^{\mathrm{c}}|}$, and (c) displays the slope of the trend-line of average diffusion distance to the Amazon at simulation time $t=0.0141$ as the time window $w$ varies, $m_w\left(d_{i,U}^{(w)}(t=0.0141)\right)$. Note that for the average non-Amazon information and proportion of non-Amazon nodes affected by diffusion, the Amazon perturbation for more recent time windows appears to diffuse faster. In (c), $t=0.0141$ is chosen due to its maximal respective inter-quartile range in $m_w(\cdot)$; the diffusion distance appears to be decreasing for the whole globe except for the Amazon itself, indicating further that perturbations over the Amazon are spreading faster. See section \ref{sec: diffusion} for more details on the diffusion simulations and metrics.
   % I have defined the initial phi such that for a connected graph, the phi value for each node will approach 1: each Amazon node is assigned a value of # nodes in graph/# amazon nodes = N/N_A = 10242/166. N_{NA}^(phi>=N_A/N)/N_{NA}: proportion of “affected” nodes, meaning their phi value is greater than N_A/N at time t. This value is chosen since it is a natural threshold between 0 and 1 to use to determine if a node is affected by the diffusion, but any value between 0 and 1 could be used. Both of these trends imply information flow can leave the Amazon and reach the rest of the network much faster. Panel c is the trendline of diffusion distance, defined d_{i,j}(t) = |phi_i(t) - phi_j(t)| and average over indices j in the Amazon. t=0.00175 was selected as the t-value to plot since it features the maximum inter-quartile range of diffusion distance trendlines. There is now a figure for all graphs (5 GTN and 5 KNN). The t value selected for each represents that which has the greatest interquartile range.
}
    \label{fig:diff-dens}
\end{figure*}

\begin{figure*}
    \centering
    \includegraphics[width=0.8\linewidth]{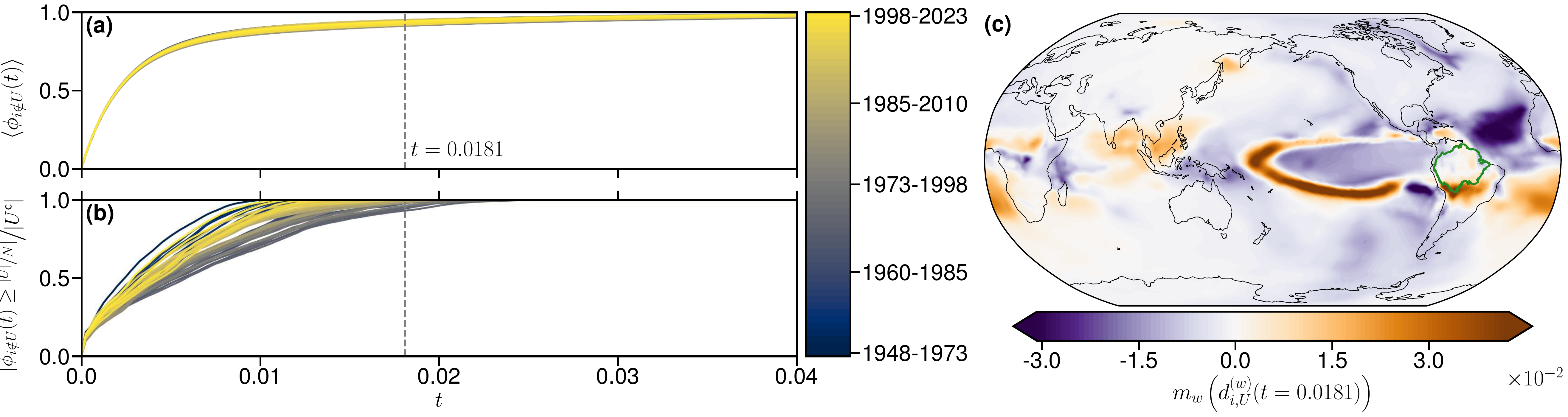}
    \caption{Diffusion simulation results for $K=500$ KNN graphs. (a) displays the average non-Amazon information, $\langle \phi_{i\notin U}(t)\rangle$, (b) displays the proportion of non-Amazon nodes affected by the diffused perturbation, $\sfrac{|\phi_{i\notin U}(t) \geq \sfrac{|U|}{N}|}{|U^{\mathrm{c}}|}$, and (c) displays the slope of the trend-line of average diffusion distance to the Amazon at simulation time $t=0.0181$ as the time window $w$ varies, $m_w\left(d_{i,U}^{(w)}(t=0.0181)\right)$. Note that for the proportion of non-Amazon nodes affected by diffusion, the Amazon perturbation for more recent time windows appears to diffuse faster, while in the average non-Amazon information the same is not apparent; this may be explained by the less significant trends in $R_{i}^{(\in)}$ seen in Fig. \ref{fig:Tr1Amazon} due to few regions of the global climate having a direct influence on the Amazon. In (c), $t=0.0181$ is chosen due to its maximal respective inter-quartile range in $m_w(\cdot)$; it appears that the regions of strongest decrease in diffusion distance are the Eastern Pacific and Western Atlantic, while the region of strongest increase in diffusion distance is the Central Pacific. This supports the trend seen in Fig.~\ref{fig:Tr1Amazon}, where the opposite trend is observed: this is explained by the fact that information is donated across outgoing edges, meaning that regions with decreasing $R_i^{(\leftarrow)}$ will be diffused a perturbation from the Amazon at a slower pace leading to increasing $d_{i,U}$, and vice versa. Due to an edge $i\rightarrow j$ in the KNN graphs meaning the presence of node $j$ in node $i$'s nearest neighbors and the fact that diffusion travels along outgoing edges, one may argue that the regions of increasing diffusion distance are losing influence over the Amazon climatic system. See section \ref{sec: diffusion} for more details on the diffusion simulations and metrics.}
    \label{fig:diff-k}
\end{figure*}

% \begin{figure*}
%     \centering
%     \includegraphics[width=0.8\linewidth]{RandomWalkTrends.pdf}
%     \caption{Random walk results for various GTN and KNN graphs. (a,b,c) display the slope of the trend-line of $\hat{p}_i(\eta=1)$ for $\rho=0.025$, $\rho=0.05$, and $\rho=0.1$ GTN graphs, respectively, while (d,e,f) display the slope of the trend-line of $\hat{p}_i(\eta=1)$ for $K=250$, $K=500$, and $K=1000$ KNN graphs, respectively. $\eta=1$ was selected due to its maximal respective inter-quartile range of $m_w(\cdot)$. Note that for the GTN graphs, the region of greatest change is the Amazon itself; $\eta=1$, so the decrease in likelihood of random walks originating in the Amazon landing in the Amazon after just a single step may imply a breaking down of the stable local continuity in the Amazonian climate system. For KNN graphs the trends are not as significant, but the edge directionality has physical meaning: random walks travel over outgoing edges, so an increase in likelihood of an Amazonian random walk remaining within the Amazon after one step indicates that the Amazon remains in its own $K$-nearest neighbors more often. This may indicate that the Amazon is becoming less influenced by local climatic phenomena.}
%     \label{fig:rwalk}
% \end{figure*}

\begin{figure*}
    \centering
    \includegraphics[width=0.8\linewidth]{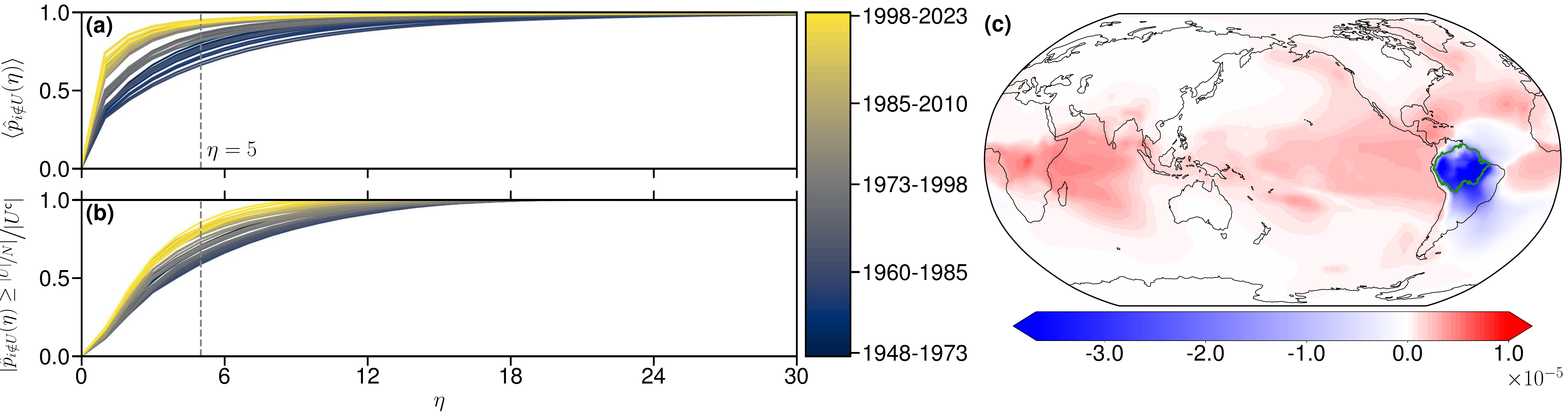}
    \caption{Random walk results for $\rho=0.025$ GTN graphs. (a) displays the average scaled (by $N$) probability of Amazon random walks of length $\eta$ reaching non-Amazon nodes, $\langle \hat{p}_{i\notin U}(\eta)\rangle$, (b) displays the proportion of non-Amazon nodes affected by the random walks, $\sfrac{|\hat{p}_{i\notin U}(\eta) \geq \sfrac{|U|}{N}|}{|U^{\mathrm{c}}|}$, and (c) displays the slope of the trend-line of scaled probability of Amazon random walks reaching node $i$ for walk length $\eta=5$ as the time window $w$ varies, $m_w\left(\hat{p}_i^{(w)}(\eta=5)\right)$. Note that for the average non-Amazon random walk probability and proportion of non-Amazon nodes affected by diffusion, the Amazon random walk perturbation for more recent time windows appears to spread faster. In (c), $\eta=5$ is chosen due to its maximal respective inter-quartile range in $m_w(\cdot)$; the region of greatest change in random walk probability is that of the Amazon itself, meaning that Amazonian random walks of length $\eta=5$ are returning to the Amazon less over successive time windows, indicating a shifting role of the Amazon in the global climate system. Inversely, the increase in random walk probability in most other regions of the world indicate growing teleconnections to the Amazon. See section \ref{sec: random-walks} for more details on the diffusion simulations and metrics.}
    \label{fig:rwalk-dens}
\end{figure*}

\begin{figure*}
    \centering
    \includegraphics[width=0.8\linewidth]{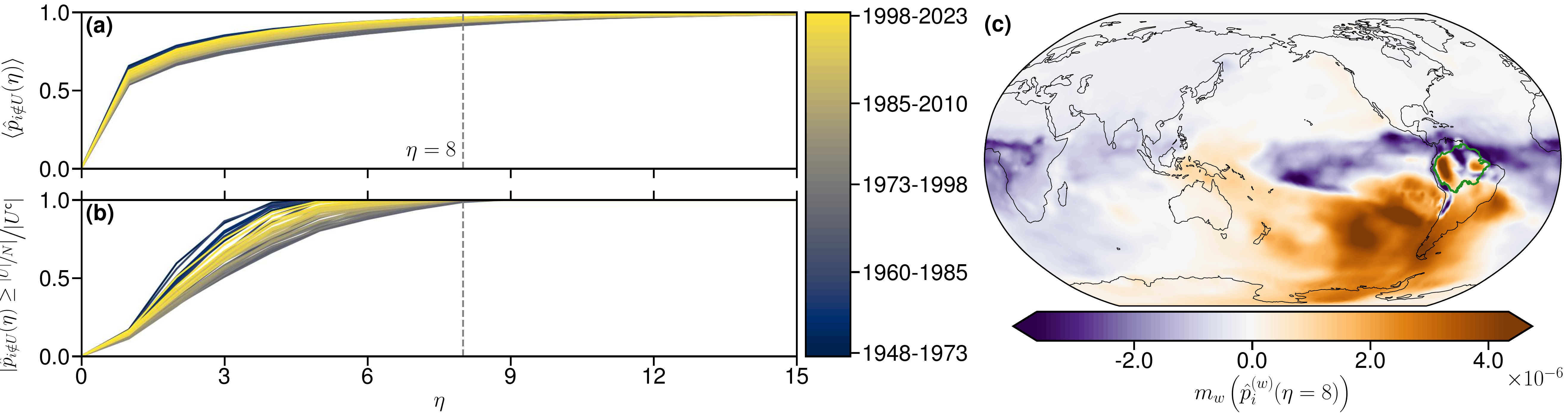}
    \caption{Random walk results for $K=500$ KNN graphs. (a) displays the average scaled (by $N$) probability of Amazon random walks of length $\eta$ reaching non-Amazon nodes, $\langle \hat{p}_{i\notin U}(\eta)\rangle$, (b) displays the proportion of non-Amazon nodes affected by the random walks, $\sfrac{|\hat{p}_{i\notin U}(\eta) \geq \sfrac{|U|}{N}|}{|U^{\mathrm{c}}|}$, and (c) displays the slope of the trend-line of scaled probability of Amazon random walks reaching node $i$ for walk length $\eta=8$ as the time window $w$ varies, $m_w\left(\hat{p}_i^{(w)}(\eta=8)\right)$. Note that for the proportion of non-Amazon nodes affected by diffusion, the Amazon random walk perturbation for more recent time windows appears to spread faster, while the lack of significant change in the average non-Amazon random walk probability may be explained in a similar fashion to Fig. \ref{fig:diff-k}. In (c), $\eta=8$ is chosen due to its maximal respective inter-quartile range in $m_w(\cdot)$; while these trends are not as significant as those seen in Fig. \ref{fig:rwalk-dens}, the directionality in the KNN graphs does have significance: since an edge $i\rightarrow j$ means $j$ is in the nearest neighbors of $i$ and the fact that random walks traverse over outgoing edges, the regions of decreasing and increasing random walk probability may indicate a decrease and increase in that region's influence of the Amazonian climate, respectively. See section \ref{sec: random-walks} for more details on the diffusion simulations and metrics.}
    \label{fig:rwalk-k}
\end{figure*}

% \begin{figure}
%     \centering
%     \includegraphics[width=\columnwidth]{RandomWalkTrends(Varydensity).pdf}
%     \caption{}
%     \label{fig:diff-dens-rwalk-trends}
% \end{figure}

% \begin{figure}
%     \centering
%     \includegraphics[width=\columnwidth]{RandomWalkTrends(Varyk).pdf}
%     \caption{}
%     \label{fig:diff-k-rwalk-rwalktrends}
% \end{figure}

\section{Results and Discussion}

Our results show an emergence of distinctive connectivity patterns between the Amazon and the rest of the climate system. While our study cannot discern the causes of these new patterns, it is nevertheless a significant scientific endeavor, allowing us to understand how the changing climate and environment are leading to the Amazon rainforest reconfiguring its connectivity patterns within the climate system. Note that the Amazon rainforest is one of the tipping elements of the global climate system, and many of these tipping elements are interconnected: triggering one of the tipping elements could have a cascading domino effect, leading to the failure of many others \cite{Steffen2018-ie}. Climate network analysis has recently provided quantitative evidence that Amazon possesses teleconnections with tipping elements. \cite{Liu2023} In contrast to the existing studies on the analysis of Amazon connectivity using climate networks, our study focuses on quantifying the new connectivity patterns that the Amazon rainforest is exhibiting within the global climate system and the possible manifestation of this new connectivity patterns in the spread of perturbation as modeled here using diffusion and random walks on networks.

We explore the new connectivity first by identifying the trends in the density of connections that Amazon has with the other regions. In Fig.~\ref{fig:Tr1Amazon}, we observe an increase in connectivity over the tropical Atlantic, the eastern Pacific, and the Indian Ocean region, some of the most dynamic regions in the global climate system. The connectivity increase with tropical Atlantic is a significant observation, as we know that the weakening of the Atlantic Meridional Overturning Circulation (AMOC) can destabilize the Amazon rainforest, leading to its dieback, as AMOC weakening can lead to variations in tropical Atlantic SST, which could change rainfall patterns over the Amazon. \cite{Steffen2018-ie,Ciemer2021}  With the Indian Ocean region, Amazon does not share a direct link. However, El Ni\~{n}o-Southern Oscillation (ENSO) is bi-directionally coupled with the larger Indian Ocean monsoon region. \cite{2010GL045932} Moreover, a recent climate networks-based study on Amazon indicates the existence of teleconnections with the Tibetan plateau. \cite{Liu2023} In Fig.~\ref{fig:Tr1Amazon} (a-c), we observe that Amazon's connectivity with the Indian Ocean and Tibetan Plateau is getting stronger. In the central Pacific, we observe two patterns: the increasing links to the region of El Ni\~{n}o pattern associated with the most significant warming, and decreasing links to the region of La Ni\~{n}a pattern associated with the most significant cooling. La Ni\~{n} phase of ENSO positively affects the Amazon as it brings wetter conditions over the Amazon. In contrast, El Ni\~{n}o brings dryer conditions over the Amazon, creating conditions for forest fires and related disasters.  

To give stronger credence to our results, we have repeated our experiments with different thresholds on the correlations and also constructed networks using distinct methodologies, GTN vs. KNN; see Fig.~\ref{fig:Tr1Amazon}. Our above findings are stable across all the repeated experiments, and we note that the interactions between some of the most dynamic regions of global climate, such as the Indian Ocean region, eastern Pacific, tropical Atlantic, and the Amazon, are getting reconfigured. Next, we explore an innovative methodology known as the Laplacian Eigenmaps to augment our above results that the connectivity between the Amazon and the global climate system is reconfiguring. In addition, this method can also identify the period in the data set when this reconfiguration started to take effect.

While employing Laplacian Eigenmaps, we observe an increase in the length of $\lVert {\bf{\xi}}_{i \in U} \rVert_{2} $ after 1992-2017 window, the 2-norm of Amazon nodes in the embedding obtained using Laplacian Eigenmaps-based projections, see Fig.~\ref{fig:LEdistances}. This increase is more pronounced in GTN with $\rho\in\{0.05, 0.1\}$ and KNN with $K\in\{500, 1000\}$, see Figs.~\ref{fig:LEdistances} (b-f). This analysis does not provide insights into the property of individual nodes or edges but provides insights into the connectivity of the Amazon within the global climate system; the first interpretation of the observations in Fig.~\ref{fig:LEdistances}  is that increase in  $\lVert {\bf{\xi}}_{i \in U} \rVert_{2} $ amounts to increase in the volume occupied by the sub-manifold of the Amazon region within the larger manifold of the global climate. Such an increase is only possible if Amazon reconfigures its connectivity with a bias towards links that allow any perturbation to (from) the Amazon to spread faster in the system, as we also observe explicitly in later results. This pattern emerges from the 1992-2017 window, the period of greatest forest loss in the Amazon.\cite{SilvaJunior2021} This observation indicates that large-scale deforestation in the Amazon could have played a role in developing the new interactions with the climate system.

Another question we explored was whether the Amazon is showing trends in connectivity that are different from areas of the same size in other parts of the globe. For this purpose, we compared the Amazon with two types of random boxes on the globe,  one lying only in the tropics (tropical random boxes) and one anywhere on the planet (overall random boxes); these boxes had the same number of nodes as in the Amazon. We observe that $\Gamma$, the connectivity ratio for the Amazon, is showing a stronger trend than the median of this quantity for the two random boxes (see Fig.~\ref{fig:met-rhp} (c,g, and k)). The Amazon is gaining outgoing links faster than the $50\%$ regions of the same size distributed in the tropics or other parts of the planet. Moreover, the geodesic length of the links between Amazon and non-Amazon nodes is also growing at a higher rate than the median of other parts of the planet (see Fig.~\ref{fig:met-rhp} (d,h, and l)). From these results, we conclude that the Amazon is gaining more long-range links faster than other regions globally. From the above observations, we hypothesize that the Amazon's local connectivity patterns are relatively stable; however, longer-range interactions in the climate system are changing; we have also observed similar patterns in Fig.~\ref{fig:Tr1Amazon}. We also note that for KNN, we do not observe consistent patterns in trends for $\Gamma$, the connectivity ratio of the Amazon, for example, whereas the trends in the connectivity ratio of the outgoing links for all the thresholds on $K$ are positive similar to GTN, for incoming links they are only positive for  $K=250$ (see Fig.~\ref{fig:met-k} (d,k, r, e, l, and s)). Also, trends in geodesic distances are all negative in KNN, contrary to our observation in GTN  (see Fig.~\ref{fig:met-k} (d,k, r, e, l, and s)), possibly due to KNN construction not being able to capture all the long-range connections.   
 
In Fig.~\ref{fig:met-k} (a,e, f, i, and j), we observe that  
in GTN, the average trends in edge betweenness, both in links internal to Amazon and going out of Amazon, is less than the median for random boxes. That is, edges involving Amazon are not gaining any strategic significance within these networks; instead, they seem to be losing some; note the slightest negative trends for most thresholds in  Fig.~\ref{fig:met-k} (a,b,e, f, i, and j). In conclusion, from GTN networks, we infer that edge betweenness trends are relatively small, close to zero; that is betweenness of the links is stable. In the KNN case, betweenness shows similar features in the edge betweenness trends but with a few exceptions, for example, $K=500$  for links within Amazon (see  Fig.~\ref{fig:met-k} (h)) and $K=1000$ for outgoing links  (see Fig.~\ref{fig:met-k} (q)). Moreover, we observe greater values of these trends in KNN and broader distribution of edge betweenness for random boxes. Given these trends' complexity and inconsistencies, we cannot draw a more significant hypothesis about changes in the edge betweenness of the Amazon for KNN.

The above analysis identified several structural changes occurring in the connectivity between the Amazon and the rest of the climate system over the last seven decades; however, this analysis does not provide quantitative insights into the possible impacts of this reconfiguring connectivity, such as how environmental or climatic perturbations from (or to) the Amazon will spread in coming years and decades. It is critical to develop such an understanding, given the increasing incidents of large-scale forest fires resulting in intercontinental smoke transport.\cite{2009GL037923,YU201373,science1092666} Furthermore, the dieback of the Amazon rainforest is now a strong possibility, and the dispersal of its cascading consequences into the global climate system stills needs to be better comprehended.\cite{Boulton2022,20211842711,pnas0804619106} In this work, we have used graph diffusion and random walks to study the transport of perturbations in the evolving climate network. To our knowledge, this is the first attempt to employ these techniques to study evolving climate networks. In the first set of simulations, we use diffusion on graphs (see section~\ref{sec: diffusion} for details), modeling the spread of perturbations from the Amazon to the rest of the globe on GTN, and found that not only perturbation can spread faster in the network but travel further, too (see Fig.~\ref{fig:diff-dens} (a-b)). An analogous analysis for KNN in Fig.~\ref{fig:diff-k} (a-b) shows an almost similar feature that, in more recent networks, perturbations can diffuse faster in the network. Moreover, in the KNN-based analysis, large parts of the central and eastern Pacific and Atlantic oceans are receiving perturbations faster from the Amazon (see Fig.~\ref{fig:diff-k} (c)). Also, for KNN networks, more complex trends coexist in some parts of the Pacific and Indian Oceans. For the GTN, in Fig.~\ref{fig:diff-dens} (c), we observe that 
the diffusion distance is decreasing for the whole globe except for the Amazon itself, indicating further that perturbations over the Amazon are spreading faster.

The diffusion discussed above is a continuous time process, and gradients in a hypothetical quantity over the graph drive it. Random walks are alternative modeling of the spreading phenomena on networks,  a discrete-time process that is not as restrictive or driven by gradients as diffusion. In the GTN networks, the random walks originating from Amazon for more recent time windows spread faster and further; see methodological details in Fig.~\ref{fig:rwalk-dens} (a-b). In Fig.~\ref{fig:rwalk-k} (a-b), a similar feature in random walks can be observed for the  KNN graphs. Moreover, in GTN networks, we note that the most significant change in the form of a decrease in random walk probability is occurring over the Amazon itself (see Fig.~\ref{fig:rwalk-dens} (c)), that is, Amazonian random walks of length $\eta = 5$ are returning to the Amazon less frequently over successive time windows, in contrast, most other regions of the globe have an increasing random walk probability. Such characteristics in the random walk can only manifest if the Amazon is building long-range connections, possibly strengthening its teleconnections. In Fig.~\ref{fig:rwalk-k} (c), we have also plotted the change in random walk probability for KNN networks; although the spatial patterns we observed do not have such clear climatological interpretation, the directionality in the KNN graphs does have significance, the regions of decreasing and increasing random walk probability may indicate a decrease and increase in that region’s influence over the Amazonian climate, respectively.

\section{Conclusion}

The Amazon rainforest is a critically endangered component of our planet's environment and climate, and it is also considered one of the tipping elements of the climate system, which can undergo irreversible changes with far-reaching consequences for the region and the globe. \cite{1739200500697x,Mataveli2022, viola2020,Liu2023,boers2017deforestation,20211842711} Therefore, identifying new, emerging features in the interactions between the Amazon and our climate system is of great significance. In this work, we analyzed the surface air temperature for the last seven decades using climate network analysis to identify and quantify changes in connectivity patterns between the Amazon rainforest region and the global climate system. After studying a collection of network properties, including trends in connectivity, betweenness, diffusion, and random walks, we conclude that the Amazon is changing its connectivity configuration, gaining longer-range connectivity with the ability to spread a perturbation faster and further.  

While carrying out the trend analysis of link densities between the Amazon and the globe, we found that the Amazon is gaining long-range links to several highly dynamic climate system components, including the western Atlantic region, critical for the future stability of AMOC. It also appears that there is a reconfiguration of connectivity between the Amazon and the South Asian monsoon region (the Indian Ocean and the Tibetan Plateau), along with the climatically critical eastern and central Pacific regions. It is important to note that AMOC and South Asian monsoons are also tipping elements. Furthermore, we observed that these new connectivity patterns allow for the faster diffusion of perturbations from the Amazon, an important observation in the context of increasing rates of forest fires that now it may be plausible for the wildfire smoke to spread faster and further.  

In this work, we also provided a few methodological innovations within the framework of climate network analysis. We showed that Laplacian Eigenmaps-based embedding could be used to study changes in the global structure of evolving climate networks; this technique could prove helpful in studying temporally evolving networks. Furthermore, we employed the method of graph diffusion and random walks on climate networks to simulate the spread of perturbations. Our results indicate that the inclusion of the technique of graph diffusion in climate network analysis could prove very fruitful in studying the impact of global warming on the spread of climatological and environmental perturbations in the global climate system.

\section*{Acknowledgement}
Authors thank Prof. Darren Narayan and Dr. Kamal Rana for many helpful discussions. 
AG, JB, and NM acknowledge the support of the National Science Foundation (NSF; Grant No. DMS-1950189, REU Site: Extremal Graph Theory and Dynamical Systems). Any opinions, findings, and conclusions or recommendations expressed in this material are those of the author(s) and do not necessarily reflect the views of the National Science Foundation.

\nocite{*}

% \bibliography{references.bib}% Produces the 

\end{document}